\documentclass{article}
\usepackage{mathptmx}
\usepackage{graphicx}
\usepackage[title]{appendix}
\usepackage{authblk}
\usepackage{setspace}
\usepackage{hyperref}

\usepackage{listings, xcolor}

\definecolor{verylightgray}{rgb}{.97,.97,.97}

\lstdefinelanguage{Solidity}{
	keywords=[1]{anonymous, assembly, assert, balance, break, call, callcode, case, catch, class, constant, continue, constructor, contract, debugger, default, delegatecall, delete, do, else, emit, event, experimental, export, external, false, finally, for, function, gas, if, implements, import, in, indexed, instanceof, interface, internal, is, length, library, log0, log1, log2, log3, log4, memory, modifier, new, payable, pragma, private, protected, public, pure, push, require, return, returns, revert, selfdestruct, send, solidity, storage, struct, suicide, super, switch, then, this, throw, transfer, true, try, typeof, using, value, view, while, with, addmod, ecrecover, keccak256, mulmod, ripemd160, sha256, sha3}, 
	keywordstyle=[1]\color{blue}\bfseries,
	keywords=[2]{address, bool, byte, bytes, bytes1, bytes2, bytes3, bytes4, bytes5, bytes6, bytes7, bytes8, bytes9, bytes10, bytes11, bytes12, bytes13, bytes14, bytes15, bytes16, bytes17, bytes18, bytes19, bytes20, bytes21, bytes22, bytes23, bytes24, bytes25, bytes26, bytes27, bytes28, bytes29, bytes30, bytes31, bytes32, enum, int, int8, int16, int24, int32, int40, int48, int56, int64, int72, int80, int88, int96, int104, int112, int120, int128, int136, int144, int152, int160, int168, int176, int184, int192, int200, int208, int216, int224, int232, int240, int248, int256, mapping, string, uint, uint8, uint16, uint24, uint32, uint40, uint48, uint56, uint64, uint72, uint80, uint88, uint96, uint104, uint112, uint120, uint128, uint136, uint144, uint152, uint160, uint168, uint176, uint184, uint192, uint200, uint208, uint216, uint224, uint232, uint240, uint248, uint256, var, void, ether, finney, szabo, wei, days, hours, minutes, seconds, weeks, years},	
	keywordstyle=[2]\color{teal}\bfseries,
	keywords=[3]{block, blockhash, coinbase, difficulty, gaslimit, number, timestamp, msg, data, gas, sender, sig, value, now, tx, gasprice, origin},	
	keywordstyle=[3]\color{violet}\bfseries,
	identifierstyle=\color{black},
	sensitive=false,
	comment=[l]{//},
	morecomment=[s]{/*}{*/},
	commentstyle=\color{gray}\ttfamily,
	stringstyle=\color{red}\ttfamily,
	morestring=[b]',
	morestring=[b]"
}

\usepackage{xcolor}
\usepackage{tikz}

\usepackage{geometry}
 \title{Transwarp-Conduit: Interoperable Blockchain Application Framework}
\author{Shidokht Hejazi-Sepehr}
\author {Ross Kitsis}
\author {Ali Sharif}
\affil{\textit{\{shidokht, ross, ali\}@aion.network}}
\affil{Aion Foundation} 
\date{January 25, 2019}
\begin{document}

\maketitle             

\begin{abstract}
Transwarp-Conduit (TWC) is a protocol for message transfers between two smart-contract enabled blockchain networks. 
Furthermore, we specify an application framework (leveraging the TWC protocol) that enables developers to define arbitrarily complex cross-blockchain applications, simply by 
\begin{enumerate}
    \item Deploying framework-compliant smart contracts and,
    \item Hosting a TWC node (daemon process).
\end{enumerate}

The TWC protocol is implementable without additional effort on part of the base blockchain protocol.\footnote{This assumes smart contract language expressivity and availability of cryptographic primitives common to most blockchains which are categorized as smart-contract enabled.}

\end{abstract}
\section{Motivation} \label{Motivation}

\subsection{Scope}
This work primarily focuses on public blockchains. Although vastly applicable, specific challenges pertaining to interoperability between private-to-private and private-to-public chains are out of the scope of this work.

\subsection{Applicability of Work}
We have observed a proliferation of public blockchains, with over 600 public blockchains in production as of end of 2018, with over 100 of these with a market capitalization of over \textdollar30M USD \footnote{Data sourced from coinmarketcap.com on December 30, 2018; used the "Not Mintable" and "Market Cap" properties of reported crypto-currencies to construct this claim.}. This diversity of well-capitalized public platforms reveals a limitation of blockchain technology: distinct blockchain constructions are required to address the vast problem-space. 

It's not obvious if there will ever exist a sufficiently-generic blockchain platform, with enough transaction capacity, ``to rule them all''\footnote{There seems to be an arms race today between several well-capitalized public blockchain organizations (Ethereum, IOHK, Parity, EOS, etc.) to invent a public blockchain platform ``to-rule-them-all'', with diverse hypotheses on scaling.}; a menagerie of specialized blockchains may become the status-quo.

In the interim, solutions are needed to interoperate between these purpose-specific and/or throughput-constrained blockchains, in order to broaden the scope and capabilities of applications that can be built on these platforms.    

 \subsection{Use Cases}
 The ability for blockchains to access foreign blockchain-state (contingent on some finality criteria) makes possible a variety of applications which were primarily the realm of centralized solutions; notable use-cases include payment-versus-delivery, payment-versus-payment, cross-chain-oracles and other general-purpose inter-chain protocols for assets and information. Two of such use cases have been of particular public interest of late:
 
 \begin{enumerate}
     \item \textbf{Decentralized exchanges:} As of this writing, all major multi-crypto exchanges (conceptually) consist of a centralized server responsible for taking and clearing orders. Interoperability infrastructure enables cross-chain exchanges to be built in a completely decentralized way: smart contracts on supported chains coordinating to take and clear orders without centralized intervention.
    
    \item \textbf {Portable assets:}  The ability to issue assets that can be transported between blockchains without violating the properties of said asset (fungibility, ownership, value, etc.) has been of great interest to create productive (smart-contract-based) economical systems and legal frameworks spanning multiple blockchains.
    
 \end{enumerate}
\section {Interoperability Definition} \label{interoperability_definition}

So far, we've invoked the term ``interoperability'' without much precision; let's precisely define the term now. 

When speaking about interoperability, we are concerned with establishing a causal relationship between events on two blockchain (isolated systems). Two\footnote{Coincidence is a third possible relationship, but it can be safely ignored since its likelihood is heuristically understood to be minuscule.} possible models exist to establish a relationships between events which need to be correlated (the task of interoperability) \cite{buterin2016chain}: causation and dependency.

Assume we have two blockchains $B_{A}$ and $B_{B}$, that must interoperate: 
\begin{enumerate}
    \item \textbf{Causation}: some internal event on one blockchain \textit{causes} a correlated internal event to occur on another blockchain. 
    \begin{enumerate}
        \item \(\hbox{\textit{\textbf{Forward causation}}: event X on $B_{A}$} \Rightarrow \hbox{event Y on $B_{B}$}\). 
        \item \(\hbox{\textit{\textbf{Reverse causation}}: event V on $B_{B}$} \Rightarrow \hbox{event W on $B_{A}$}\). 
    \end{enumerate}
    \item \textbf{Dependency}: some external event \textit{causes} a pair of internal events to occur on both $B_{A}$ and $B_{B}$ (i.e. both blockchain events correlated by the same external event). 
        \begin{enumerate}
        \item \(\hbox{E.g. external event Z} \Rightarrow \hbox{(event X on $B_{B}$ AND event Y on $B_{A}$)}\). 
    \end{enumerate}
\end{enumerate}

From the perspective of certain applications, a dependency relationship may be sufficiently expressive (e.g. an oracle publishing the same data stream to multiple blockchains). But most interoperability relationships, particularly the ones described in section \ref{Motivation}, require the structure of causation for their articulation. 
For the purposes of this paper, when we say: \\
\centerline{``Two blockchains interoperate'',}
we mean:
\begin{center}
    ``There exists a mechanism to establish causation between transactions sealed onto distinct blockchains\footnote{The blockchains have different genesis blocks and share no history.}''.
\end{center}

\section{Design-Space} \label{design_space}
This section outlines the trade-offs available to the authors with respect to available interoperability architectures.

A multi-signature notary scheme was the implementation-strategy of choice for the TWC protocol, since it provided a reasonable trade-off between decentralization (by opting into a well-defined trust model), ease-of-implementation, system reliability and on-chain data-amplification. 

Three families of interoperability mechanisms were identified for analysis: 
\begin{enumerate}
    \item Time-locks (hash locks)
    \item Relaying (validate-one-chain-inside-another)
    \item Notary schemes
\end{enumerate}

An exposition of each of these methods is provided to highlight key design trade-offs; a more nuanced discussion of these mechanisms is available in \cite{buterin2016chain} and \cite{lerner2016drivechains}.

\subsection{Time Locks (Hash Locking)}
These systems comprise of some time-locked operation across two chains, which triggers on the revelation of the pre-image of a one-way function. We illustrate this concept with the well-known example of hash locking (a.k.a. ``atomic swaps''):

Assume entities X and Y need to swap some asset that exists on two distinct chains ($B_{1}$ and $B_{2}$). Assume some asset $C_{1}$ is owned by X on $B_{1}$ and asset $C_{2}$ is owned by Y on $B_{2}$. Furthermore, users X and Y have accounts on both blockchains ($X_{B1}$, $X_{B2}$, $Y_{B1}$, $Y_{B2}$). There exists some well-defined time interval T. Also assume some ideal hash function H. 
\begin{enumerate}
    \item X generates a secret S and generates a hash H(S) and sends H(S) to Y.
    \item $X_{B1}$ locks asset $C_{1}$ on $B_{1}$ with the following conditions: 
    \begin{enumerate}
        \item If S is revealed within time 2T, $C_{1}$ is transferred to $Y_{B1}$, 
        \item Else, $C_{1}$ is transferred back to $X_{B1}$.
    \end{enumerate}
    \item After Y sees X lock asset $C_{1}$ on $B_{1}$, Y uses their account on $B_{2}$ ($Y_{B2}$) to lock asset $C_{2}$ with the following conditions:
    \begin{enumerate}
        \item If S is revealed within time T, $C_{1}$ is transferred to $X_{B2}$,
        \item Else, $C_{2}$ is transferred back to $Y_{B2}$.  
    \end{enumerate}
    \item X reveals secret S on $B_{2}$ to claim $C_{2}$. 
    \item Y observes S on $B_{2}$ and publishes it on $B_{1}$ to claim $C_{1}$. 
\end{enumerate}

\subsubsection{Analysis}

Several implementations and discussions of this scheme exist \cite{herlihy2018atomic, interledger2018hashed}; the discussion on this topic here is limited in so far as it illustrates the following relevant facts about this system: 
\begin{enumerate}
    \item The protocol is only capable of defining a \emph{dependency} relationship between events on the two blockchains ($B_{1}$ and $B_{2}$); in our illustration,  the asset transfer events on $B_{1}$ and $B_{2}$ are dependent on the revelation of the pre-image of the hash. 
    \item This protocol requires a high degree of interactivity; each user (X and Y) is required to interact with both blockchains at least once, in addition to off-chain (real-time) requirements of observing and reacting to events on a blockchain. 
\end{enumerate}

\subsection{Relaying}
This mechanism relies on the destination chain $B_{D}$, validating the consensus rules of the source chain $B_{S}$ (making the destination chain effectively a ``light-client'' of the source chain). An event ($E_{D}$) on the $B_{D}$ can now depend on an event ($E_{S}$) on $B_{S}$ through light-client verification techniques available for the $B_{S}$ (e.g. providing a Merkle-proof for $E_{S}$'s membership in a block on $B_{S}$). Let's illustrate this with an example: \\
\linebreak 
Assume two blockchains, source chain ($B_{S}$) and destination chain ($B_{D}$), which both have the following properties: 
\begin{enumerate}
    \item They have the notion of the block headers, which is a data structure whose hash can uniquely identify a block \cite{wood2014ethereum}. 
    \item The block headers are some finite, well-defined size that is much smaller than the full block contents.  
    \item They collect all transfers included in a block into a Merkle tree, and then store the root of that tree in the block header.
    \item The consensus rules can be validated completely by examining the block headers.
\end{enumerate}

Now let's construct a smart contract $R_{D}$ that has functions that: 
\begin{enumerate}
    \item Accepts block headers from the source blockchain $B_{S}$, validates the consensus rules of the source blockchain and maintains a ``light-replica'' of $B_{S}$ in the form of a chain of block headers.
    \item Accepts a transaction hash, a Merkle proof and a source block number, and validates that the transaction was sealed in the source chain. 
\end{enumerate}

Now one could imagine constructing a smart contract on $B_{D}$ that leverages our relay contract $R_{D}$ to accept proof-of-inclusion of some transaction that occurred on $B_{S}$ before executing its logic. This effectively allows one to establish a \textit{forward causation} relationship (section \ref{interoperability_definition}) between $B_{S}$ and $B_{D}$.

\subsubsection{Analysis}

One powerful property of this system is that it's completely trustless:
\begin{enumerate}
    \item The relay contract $R_{D}$ enforces the source chain's consensus rules in a trustless smart contract (publicly auditable)
    \item Anyone can run a process to relay (hence the name) the source chain ($B_{S}$) block headers to $R_{D}$. Notice that liveness of the whole system is dependent on at-least one relaying entity being online.  
\end{enumerate}

But there are some problems with this system: 
\begin{enumerate}
    \item The amount of data required on the destination chain is proportional to the size of all the source chain's block headers since its origin block. This can be problematic due to block size limitations on the destination chain. This can be further exacerbated by an impedance mismatch between the chains; if the source chain's block time is much faster than the destination blockchain, then the size of the source block headers per destination block can become a significant proportion of the block limit. Several solutions to this problem have been proposed in \cite{meckler2018coda} and \cite{kiayias2017non}, and are actively under investigation. 
    \item This scheme requires the re-implementation of the consensus algorithm in the smart-contract language of the destination chain. In a world with diverse blockchains, there are several problems with this approach: 
    \begin{enumerate}
        \item Blockchain platforms use a diverse range of cryptographic primitives to validate their consensus algorithms; most smart contract languages currently have limited support for cryptographic primitives other than the ones used natively.
        \item Even if all the required primitives to validate the diversity of consensus algorithm implementations were universally available, re-implementing the same algorithm across a non-trivial number of platforms carries significant operational risk and costs.  
    \end{enumerate}
\end{enumerate}

In the opinion of the authors, exclusive usage of relaying to power a flexible interoperability solution requires further research and system design.
\subsection{Notary Schemes}
In this scheme, some external (notary) entity is allowed to make claims on one blockchain about what happened on another blockchain. For this to work, the blockchain must trust the notary entity to honestly make claims about some other external blockchain. 

Since a primary design goal of blockchains is decentralization, a combination of several strategies can be used to mitigate fragilities which arise from trusting a single external entity:
\begin{enumerate}
    \item Use a set of well-known entities, who are disincentivized to deviate from protocol, either via reputation-at-stake mechanism or bonded financial resources. 
    \item Force the notary entities to put up financial stake, which they can stand to lose if they misbehave. 
    \item Have a plurality (some appropriately large number) of signatories, who need to agree on the same value, either interactively via a consensus algorithm, or non-interactively via a multi-signature scheme. 
\end{enumerate}

\subsubsection{Analysis}

This scheme offers the most versatility in establishing causation relationships between transactions on two blockchains with diverse structures and capabilities; since the relay and signatory processes can be implemented in any general-purpose language, they serve to “translate” between two blockchains. Let's illustrate this idea with an example: 

Assume two distinct, interoperating blockchains \(\hbox{$B_{1}$} \rightarrow \hbox{$B_{2}$}\) use different asymmetric cryptographic systems ($\Omega_{1}$, $\Omega_{2}$) and one-way functions ($H_{1}$, $H_{2}$). Since the signatory processes are implemented on some general-purpose computation platform (e.g. linux on AMD64), there is no restriction on the crypto-system which can be implemented therein. The smart contract implementation on $B_{2}$ only needs to understand $\Omega_{2}$; the signatory can cryptographically attest to its observations of $B_{1}$ using $\Omega_{2}$.

\subsection{Architecture Selection}
At the time of authorship of the TWC framework report, a notary scheme with a well-defined trust model was considered the best path to a production interoperability solution in the very short term; absolute decentralization was traded-off against its flexibility in enabling a uniform interoperability solution across a diversity of blockchains.

Hash locking was precluded from consideration on part of its inability to express causality relationships between chains.

Although the relaying scheme offers very attractive decentralization properties, write amplification and limitations of today's smart contract platforms make a uniform implementation across a diversity of blockchains infeasible in the short term. As capabilities of smart contract platforms improve, this will become a more compelling architecture for interoperability. 

\section{System Overview} \label{Overview}

\subsection{Design Objectives}

The following objectives were adhered-to when making system-level design decisions:

\begin{enumerate}
    \item \textbf{Keep-it-simple}: a simpler protocol is heuristically less error-prone and easier to reason about.
    \item \textbf{Favor safety}: the protocol should favor safety (primarily when trade-off is against liveness).
    \item \textbf{Minimize interactivity}: the protocol should require minimal interactivity from the end user; it should make no assumptions about the online-ness of the user of the protocol, beyond those made by the blockchain where a transaction originates.
\end{enumerate}

\subsection{Protocol Description}

This section provides an overview of the (a detailed description is available in  section \ref{Protocol_Spec}).\\

For the purposes of this discussion, this protocol is modelled as a message passing mechanism between a source blockchain and a destination blockchain. Note that the protocol is bi-directional; the unidirectional case is presented for expository purposes. The reverse information flow is simply a dual (symmetric in the other direction).

The simple construction is as follows:
\begin{enumerate}
    \item A message M gets sealed into block $N_{S}$ on a source blockchain $B_{S}$.
    \item Wait for $F_{S}$ blocks to elapse on source blockchain $B_{S}$ ($F_{S}$ being the acceptable finality on source chain).
    \item At some time $\varepsilon$ after block number $N_{S}$ + $F_{S}$ is sealed on $B_{S}$, message M gets sealed into block $N_{D}$ onto the destination blockchain $B_{D}$. (Note that $\varepsilon$ models the time required for the ``message pasting action'' to take place). 
    \item Wait for $F_{D}$ blocks to elapse on destination chain $B_{D}$ ($F_{D}$ being the acceptable finality on the destination chain).
\end{enumerate}

The above construction contained nothing about how the system works, merely what it does. In order to describe how the protocol works, we first outline the actors involved and their responsibilities; we then summarize the interactions between them.
\begin{enumerate}
    \item Blockchain ($B_{S}$, $B_{D}$): source and destination blockchains respectively, for some message M.
    \item Signatories ($S_{n}$, n \textgreater \ 0): passive entities, which upon the request of the relay, make independent observations on source blockchain ($B_{S}$) and attest to the
    \begin{enumerate}
        \item History (when M was sealed onto $B_{S}$) and,
        \item Contents of M. 
    \end{enumerate}
    \item Relay (R): coordinates the flow of the message from source to destination chain:
    \begin{enumerate}
        \item Actively monitors $B_{S}$ for any messages addressed to be replicated on $B_{D}$. 
        \item Collect signatures on the history and content of M from 2/3 of $S_{n}$.
        \item Send a transaction to $B_{D}$ containing M. 
    \end{enumerate}
\end{enumerate}

We need one more bit of exposition, before we're ready reveal the protocol in its entirety. So far we have been suspiciously vague about the nature of the message M, so let's define that now. 

The system assumption we made at the outset of this document is that the blockchains are being ``bridged'' must: 
\begin{enumerate}
    \item Have a sufficiently expressive smart contract language to define arbitrary user-defined functions, and
    \item Must implement language primitives necessary to validate signature (the signature algorithm can be anything).  
\end{enumerate}

The message M at the source blockchain is a transaction containing message M, that is the successful execution of the ``egress'' function \textit{requestTransfer(message)} on a smart contract we refer to as the ``Adapter'' well known to Relay (R). The implementation of the \textit{requestTransfer(message)} function must adhere to the specs defined in section \ref{Protocol_Spec}.

The message M at the destination blockchain is a transaction that consumes the message generated from the egress function of the Adapter contract deployed on the source blockchain. In particular, the destination smart contract must implement the ``ingress'' function \textit{processTransfer(message, signatures)} as defined in section \ref{Protocol_Spec}. 

In particular, the message M is composed of two features: 
\begin{enumerate}
    \item History: when M was sealed onto $B_{S}$
    \item Contents: a binary payload encoding the desired function call and arguments on the destination chain (as defined in section \ref{Protocol_Spec})
\end{enumerate}

\begin{figure}[h!]
  \includegraphics[width=\linewidth]{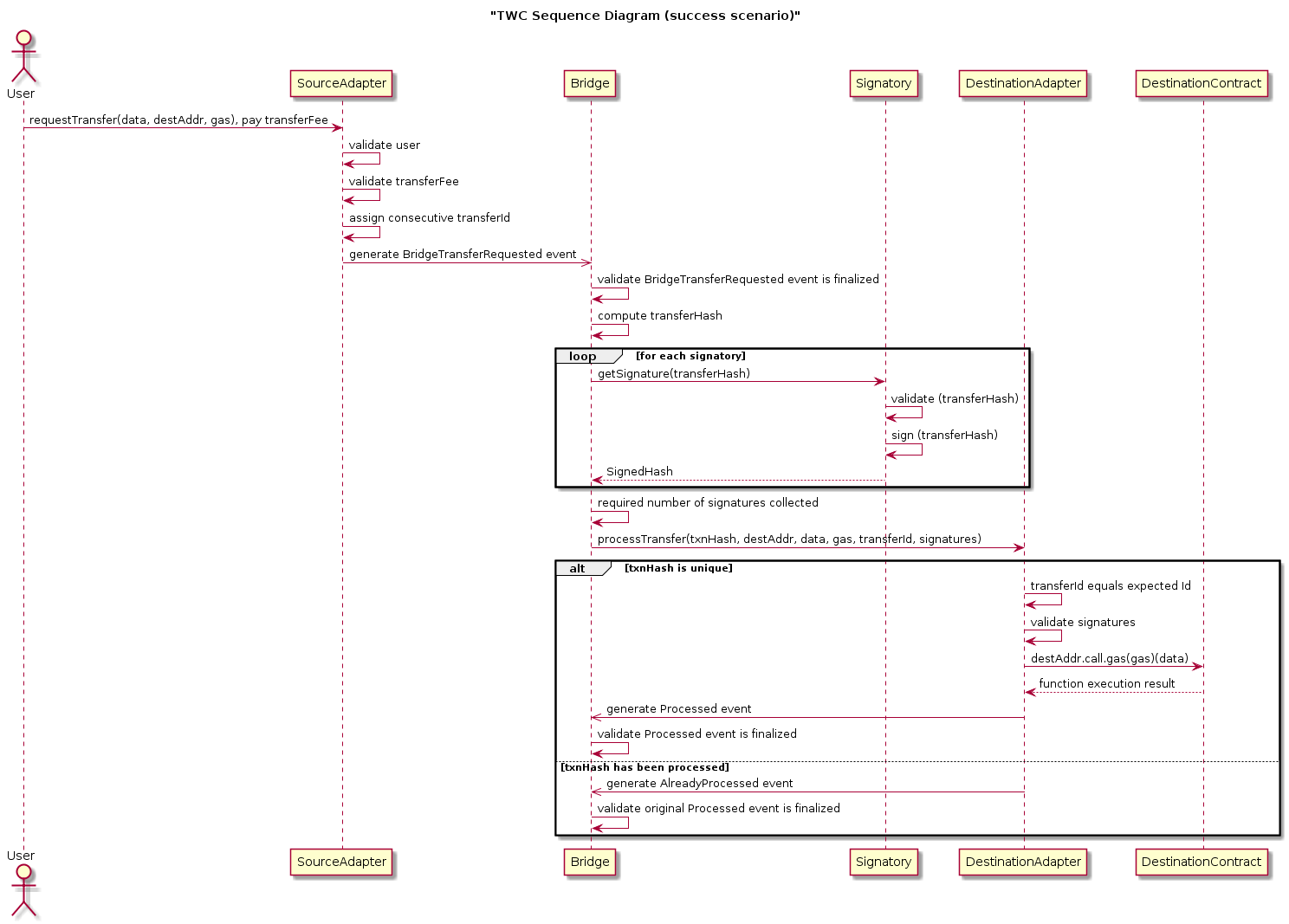}
  \caption{Summarizes the ``happy path'' of the protocol in its entirety.}
  \label{Figure 1}
\end{figure}

%
\subsection{Trust Model}
The trust model is implemented entirely in the destination smart contract. The destination smart contract must implement the following rules: 
\begin{enumerate}
    \item The smart contract must have access to a pre-defined list of $N_{sig}$ Signatories (public keys) whose signatures on the history and contents of M it implicitly trusts. 
    \item The smart contract must define a Relay (address) which is the only entity allowed to post ingress messages to the contract.
    \item The ingress function \textit{processTransfer(message, signatures)} shall implement the following rules: 
    \begin{enumerate}
        \item At least 2/3 of $N_{sig}$ signatures are required before message is processed.
        \item Only the Relay is allowed invoke this function.
    \end{enumerate}
\end{enumerate}

As it's obvious from the construction, the entire trust model is encoded in the destination smart contract. A quorum of signatories is responsible for attestation of the source message; i.e. as long as a two-thirds majority of the signatories agree that message M was observed on the source chain, the destination chain accepts that as fact and proceeds to process M. 

Furthermore, in this model, the Relay entity (R) is trustless; the dependency in the trust model exists only as a first-order implementation of a "paranoid fencing" system a service-provider might implement to ensure contract invocation by trusted accounts to prevent public exploitation of bugs discovered in the production smart contract. This framework is not opinionated in the incentive mechanisms for the Relay to continue providing service; these incentives can be implemented in an application-by-application basis (at the smart contract level). The cost of the transfer is borne by the Relay entity, which serves as a built-in denial-of-service deterrent.

\subsection{Safety and Liveness}
We provide a crude sketch for the safety and liveness arguments for this system. 

\subsubsection{Safety Sketch}
For the purposes of this sketch, we use the heuristic definition of safety: ``nothing bad will ever happen''. The safety of this system is dependent on the set of signatories and the smart contract implementation. The safety of the system can be guaranteed if: 
\begin{enumerate}
    \item At-least two-thirds (2/3) of the signatories make faithful observations on the source blockchain $B_{S}$'s mainchain, and
    \item If all protocol constraints defined in section \ref{Protocol_Spec} are implemented correctly in the ingress function on the Adapter contract on the destination chain ($B_{D}$).
\end{enumerate}

\subsubsection{Liveness Sketch}
For the purposes of this sketch, we use the heuristic definition of liveness: ``something good will eventually happen''. The liveness of the system is dependent on the set of signatories, the relay and the blockchains involved. The liveness of system can be guaranteed if:  
\begin{enumerate}
    \item The liveness of the blockchains being ``bridged'' is not violated, 
    \item At-least two-thirds (2/3) of the signatories respond to requests for signatures from the Relay within some pre-defined time (synchrony),
    \item Some entity is providing a Relay service with liveness guarantees.   
\end{enumerate}

Notice that the relay is only responsible for liveness; it needs to be always online and maintain at least one connection to each of the blockchain clients being ``bridged''. If the relay is byzantine and the signatories are not, the system comes to the halt (i.e. messages are not replicated on the destination chain) and no malicious messages get delivered to the destination chain. 

A more in-depth discussion of the failure and liveness of the system is available in section \ref{Failure_Modes}.

\section{Protocol Specification} \label{Protocol_Spec}
This section provides a high-level description of the TWC operation and usage. The TWC is split into 4 components:
\begin{enumerate}
    \item Developer defined smart contract
    \item Adapter contract
    \item TWC bridge node
    \item TWC signatories
\end{enumerate}
\subsection{Developer Defined Smart Contract}
This smart contract is any contract or application defined by the developer; the TWC places no restrictions on the functionality or implementation of the contract. Initial use cases are expected to be cross chain asset movement, however, the TWC is a generic protocol allowing for the transfer of any arbitrary data; not just that of representing assets. 

Any interoperable smart contract must implement a function to form the ABI encoded data and call \textit{requestTransfer} of its corresponding Adapter.
Optionally, the source smart contract should contain developer defined external remediation functionality. The security model section \ref{Security_Model} provides details for scenarios which may cause inconsistencies between the source and destination chains.
\subsection{Adapter Contract}
The Adapter contract is the primary interface between the TWC node and the developer defined smart contract. At the very least, the Adapter should be initialized by providing a valid set of signatories, Relayer account address, minimum number of required signatures and cross-chain transfer fee. Additionally, an authorized set of senders should be provided if the request initiators are to be validated. The core functionality of the Adapter is to:
\begin{enumerate}
    \item Accept incoming messages (encoded function calls) from developer smart contracts or applications through \textit{requestTransfer}. 
    \begin{enumerate}
         \item Assign a nonce to each transfer and charge usage fees as specified by the owner. Optionally, it may also verify senders, preventing unauthorized users from initiating cross chain transfers.
         \item Emit outgoing message events to be picked up by TWC (\textit{BridgeTransferRequested})
    \end{enumerate}
       \item Accept incoming messages from the TWC node (identified by the Relayer account) through \textit{processTransfer}. 
    \begin{enumerate}
    \item Verify transfer signatures against the current set of signatories and threshold.
    \item Maintain a list of processed messages and reject duplicate messages.
    \item Track nonces of previously sent messages and reject out-of-order transfers.
    \item Call developer smart contracts.
    \item Emit events indicating the transfer status (\textit{Processed/AlreadyProcessed)}.
    \end{enumerate}
\end{enumerate}

Detailed functionality and source code of Adapter contracts is provided in Appendix \ref{contract_code}.
\subsection{TWC Bridge Node}
The TWC bridge node is the primary component connecting two distinct blockchains and it is responsible for the following:
\begin{enumerate}
    \item Monitor connecting chain status
    \item Monitor Adapter smart contract events
    \item Process transfer requests
    \begin{itemize}
        \item Extract transfer data from source chain events
        \item Maintain transfer order based on a sequential nonce
        \item Request signatures
        \item Send transfers to the destination chain
        \item Monitor transfer status
    \end{itemize}
\end{enumerate}
\subsubsection{Monitoring Connecting Chain Status}
The TWC node must monitor the status of both $B_{S}$ and $B_{D}$. Monitoring the status of a blockchain will vary depending on chain properties, thus specifics are left to the implementation. In a general sense, chain status must encompass at least \textit{liveliness} and \textit{reorganization}. Liveliness refers to the generation of new blocks, while reorganization events lead to a chain's history being re-written or switching to a different branch.   

In addition to monitoring chain status the TWC must also respond when detecting incorrect behavior. Responses will vary by implementation however there are 3 strategies that may be followed:
\begin{enumerate}
    \item Pause operation until operator intervention
    \item Retry operations as needed
    \item Continue as normal
\end{enumerate}
\subsubsection{Monitoring Adapter Smart Contract Events}
In order to detect transfer requests and responses, TWC node must monitor each connecting blockchain for events generated by the Adapter contracts. 
From the source Adapter the TWC must monitor the following events:
\begin{itemize}
    \item \textbf{BridgeTransferRequested:} Specifies a new transfer
\end{itemize}
From the destination Adapter the TWC must monitor the following events:
\begin{itemize}
    \item \textbf{Processed:} Indicates a transfer has been processed. 
    \item \textbf{AlreadyProcessed:} Indicates a submitted transfer was previously processed and has been resubmitted. 
\end{itemize}
\subsubsection{Processing Transfer Requests}
Processing transfer requests is the primary role of the TWC node and it is performed in the following steps:
\begin{enumerate}
    \item Detect a \textit{BridgeTransferRequested} event from $B_{S}$.
    \item Extract transfer details from the transfer request.
    \item Create a signing request.
    \item Send signing request to all $S_{n}$ for verification and signatures.
    \item Collect signatures and create a transfer transaction.
    \item Submit the transfer to $B_{D}$. 
    \item Monitor $B_{D}$ to ensure the transfer transaction is successful.
    \item Listen for \textit{Processed} or \textit{AlreadyProcessed} events from the Adapter smart contract. 
\end{enumerate}
In addition to processing transfer requests, TWC nodes must also provide a guarantee on transaction ordering i.e. the TWC must process transactions in the same order as they occur on the source chain. 
\subsection{TWC Signatories}
The primary function of the signatories is to process signature requests, validate transactions and finally sign valid transfers. Transfer sign requests contain the minimal amount of data needed for verification:
\begin{itemize}
    \item Source chain block number
    \item Source chain block hash
    \item Source chain transaction hash
    \item Source chain transaction data hash
\end{itemize}
Using the block number, block hash and transaction hash signatories must: 
\begin{enumerate}
    \item Fetch the specified block in order to ensure their view of the chain is the same as that of the TWC node.
    \item Fetch the specified transaction, take a hash of transaction data and compare it to the provided transaction data hash. 
    \item Given the provided transaction data hash and calculated data hash are equal, the signatories will sign the transaction data hash using their private key. 
\end{enumerate}
TWC nodes collect these signatures in order to attest to an event occurring on the source chain.
\section{Failure Modes} \label{Failure_Modes}
Proof-of-Work (PoW) blockchains as a standalone data structure and application share several common attack vectors and behaviors, these include:  
\begin{enumerate}
    \item 51\% attacks leading to reversion of transactions. 
    \item 51\% attacks leading to transaction censorship.
    \item 51\% attacks leading to the creation of an invalid chain which is detected by full nodes. However, light nodes may accept the invalid chains. 
    \item Soft forks.
    \item Hard forks (bulk of the user bases switches to the new chain). 
    \item Network splitting into multiple fragments. 
    \item Consistency favoring chains halting due to a network issue.
\end{enumerate}

Any one of these events may cause serious issues for a single chain; interoperability between chains amplifies the effects of each of these events. Cross-chain application developers must be cognizant of chain failure events and build chain failure checks into their applications. Even when handling chain events within an application developers and operators must be ready with off-chain solutions; certain cases detailed in the following section may not be detectable until it is too late to stop the invalid cross chain transaction. These cases will require an off-chain solution to correct the invalid state introduced on the destination chain.  

During exceptional circumstances chains may become compromised prompting emergency responses such as the DAO or bitcoin overflow fork. The TWC significantly complicates these types of forks, once a message has been sent across the TWC, a fork on the source chain cannot reverse transactions on the destination chain. These types of events have the potential to lead to attacks such as double spending of transferred assets on the destination chain. While these types of events are extraordinarily rare, developers and operators should be aware of the impact of these types of events.  

Given the current state of blockchain technology; without radically changing blockchain implementations a fully decentralized, failure proof solution is impossible to achieve. Some sort of trade-off between risk, centralization and interactivity must be made by interoperability solutions, broadly these may be defined as: 
\begin{itemize}
    \item \textbf{Centralization:} The level of control a single actor may exert over a system; given a fully centralized system a single actor may fully control cross chain functionality while in a fully decentralized system no one actor may control cross chain functions. 
    \item \textbf{Risk:} The level of risk operators are willing to accept. A higher level of risk may simplify operation and increase throughput, however, chain events may be exploited to create invalid states on the destination chain. 
    \item \textbf{Interactivity:} The level of interaction required by users in order to validate bridge operation. As the level of interaction increases, users are always expected to be online and monitor the operation; should users be offline invalid transactions may be submitted and accepted on the destination chain.
\end{itemize}

The TWC and notary-based schemes in general choose to trade centralization for interactivity. Very little interactivity beyond initiating the transfer is required from the user. However, to achieve the low level of interactivity the TWC must inherently trust its set of signatories to validate and attest to the validity of transfers.  

The remainder of the section focuses more specifically on the security model and vulnerabilities in the described notary based TWC bridging solution. 
\subsection{Security Model} \label{Security_Model}

The security model of the TWC is provided by the signatories making faithful observations of the source chain as well as the Adapter contracts. Given these constraints, no single compromised system component may compromise the entire system. 

\subsubsection{Risk and Impact}
Before exploring the TWC security model a ranking is assigned to the risk of exploiting an attack vector and the potential impact a compromised party may have on the TWC. Impact is examined from the view of contracts entering invalid states. Propagation of invalid transactions to the destination chain exponentially increases the damaging impact of an attack.

\textbf{Risk levels:}
\begin{itemize}
    \item Low: Attack requires multiple independent parties to be simultaneously compromised.
    \item Medium: Attack requires a single party to be fully compromised. 
    \item High: Attack may proceed without any parties fully compromised. 
\end{itemize}

\textbf{Impact levels:}
\begin{itemize}
    \item Low: It may temporarily shut down a bridge and cause inconvenience. No data will be corrupted on either $B_{S}$ or $B_{D}$ and no external remediation is required.
    \item Medium: One or more bridges are totally censored and unable to transfer transactions. No corruption of contract states; minor remedial actions may be required mainly in configuration changes.  
    \item  High: $B_{S}$ or $B_{D}$ contract state corrupted. Possible double spend attacks, theft or complete contract state invalidation. Remediation will likely be required by the contract owner (if possible). 
\end{itemize}
\subsection{Single Party Compromise} \label{Single_Party_Compromise}
\subsubsection{Source Chain Infrastructure (Risk: Low, Impact: Low-High)}
A compromised source chain is defined as at least 2/3 of connections to $B_{S}$ providing faulty data, causing the bridge to detect and process incorrect data. The impact of such an attack may be broken up into two scenarios:
\begin{itemize}
    \item Individual nodes return incorrect data; however, the chain itself is intact and signatories follow the original chain. In this scenario, there will be no impact on the destination chain; signatories will reject the invalid transactions and they will not be submitted to the destination chain.
    \item Signatory chain connections may also return invalid data. This may cause the destination contract to reach an invalid state. Such a scenario is equivalent to multiple parties or the entire source chain being compromised, and will be explored further in section \ref{Multi-Party_Compromise}. 
\end{itemize}
\subsubsection{Destination Chain Infrastructure (Risk: Low, Impact: Low)}
A compromised destination chain is defined as at least 2/3 of connections to $B_{D}$ providing faulty data. The impact of this attack is quite low as actions are triggered by the source chain. At worst an attacker may be able to stall the bridge or cause it to shut down due to inconsistent messages. Attacks of this nature may be mitigated by moving to alternative chain connections where bridge operation may resume. 
\subsubsection{Adapter Contract (Risk: Medium, Impact: High)} \label{Adapter_compromise}
The Adapter contract is a critical piece of the TWC, thus compromising it on either the source or destination chain may have quite serious consequences. In order to compromise an Adapter, an attacker needs to gain access to its owner's private key. Adapter owner account can modify important properties such as authorized senders, signatory public keys and Relayer address. Combining a subset of these changes will allow an attacker to corrupt the state of user contracts on either $B_{S}$ or $B_{D}$. Attacks could be performed on source or destination functionality.

One possible attack on the source side would be to modify the authorized sender list to allow an attacker to use their own contract to generate transfer requests in the place of a user's contract. This will allow attackers to bypass a user's defined logic by substituting their own. The TWC node and signatories have no way of distinguishing the original source of the transaction, thus the transfer will appear valid. The message will be sent to the destination chain and accepted by the destination Adapter; setting an invalid state on the destination contract. 

A possible attack on the destination would modify the set of signatory public keys as well as the Relayer account; giving an attacker the ability to send any message to the destination contract, completely cutting out the TWC node and signatories. Invalid messages would have the ability to set an invalid state in the destination contract. 

Attacks of this nature are less TWC specific and are more a generic class of attacks on all smart contracts; once an owner's private key has been compromised it is inevitable that the attacker will gain a large amount of control over the contract; which he may exploit in malicious ways. The extent of the damage an attacker could do in the context of TWC is difficult to quantify as different contracts will have different levels of access.

The TWC protocol spec regarding Adapters mandates each change in the contract to generate a corresponding event. While the spec does not specify monitoring and usage of those events, a TWC node or an external process could be set to monitor them. If an unexpected change is detected, the monitoring process could take emergency measures such as shutting down the TWC to ensure those changes do not allow cross-chain attacks. 
\subsubsection{Bridge Node (Risk: Medium, Impact: Low)} \label{Bridge_Compromise}
A compromised bridge may perform 4 types of malicious behaviors:
\begin{enumerate}
    \item \textbf{Request signatories to sign invalid transactions:} 
    Signatories independently verify each signing request; without locating and verifying the corresponding transaction on the source chain, a signatory will not sign the request. Therefore, sending invalid transactions for signatures will not impact bridge, source or destination chains. 
    
    Stemming from this type of attack is attempting to flood signatories with invalid signature requests. Attacks of this nature may be mitigated by rate limiting each connection. A malicious bridge may max out its own connection, however, this will not negatively affect other bridges. 
    \item \textbf{Send invalid transactions to the destination chain:} 
    The destination chain Adapter will verify each signature based on its configuration. As the bridge is unable to falsify signatures, the destination chain will detect invalid transactions and reject them. 
    \item \textbf{Censor transactions:} 
    All events are assigned a nonce (\textit{transferId}) by the source Adapter contract allowing the destination Adapter to detect missing transactions and reject out of order nonces. Attempting to censor a transaction will lead to a bridge stalling as the destination chain will not process transactions until the expected nonce is submitted. While this will stop the following legitimate transactions, we believe it is an acceptable trade-off since it will immediately alert the bridge operator and users of incorrect behavior, allowing them to act accordingly. 
    \item \textbf{Replay transactions:} 
    A compromised bridge may store signed transactions and replay them later, without having to request signatures from signatories. Destination Adapter maintains a history of all processed transactions; hence, replayed transactions will be rejected by the Adapter. 
\end{enumerate}
\subsubsection{Signatories (Risk: Low, Impact: Medium)}
A compromised signatory set is defined as at least 2/3 of $S_{n}$ censoring transactions; either by refusing to sign validation requests or by providing incorrect signatures.
\begin{itemize}
    \item \textbf{Refusal to sign:} In this scenario signatories will simply stop replying to bridge requests, allowing them to time out. A bridge will not be able to distinguish this case from a network error and will respond in the same fashion. Without noting specific implementation details the only option a bridge will have is to re-try sending the message, either indefinitely or for some period before giving up on the request. The state on the destination chain will not be corrupted. 
    \item \textbf{Incorrect signatures:} In this scenario signatories will reply to validation requests, however, the replies will contain invalid signatures. Bridge performs cursory checks on signatures, but since it is unaware of the specifics of a signature, it is unable to verify it. If the signatures pass all cursory checks, the bridge will forward the transfer to the destination chain. This transfer will fail the signature validation step in the Adapter contract and the state will not be modified.  
\end{itemize}

In both scenarios the state of the destination contract will be protected by either the bridge or the Adapter contract. Messages on the source chain may be stuck in a pending state while signatory functionality is restored; however, no messages will ever be lost. While inconvenient, bridge operators may switch the signatory set to begin processing backlogged messages or simply wait till the signatory behavior is corrected.
\subsubsection{Operator (Risk: Low/Medium, Impact: Medium/High)}
An operator is a special actor within the TWC owning a significant share of the TWC infrastructure, as such the operator has a great deal of control over the TWC. Due to the high level of access in the user defined contract, the operator is generally seen as controlling the TWC node and Adapter contracts; in most cases they will also control the user contract however this is not strictly required. Consequently, an operator will have access to the following:
\begin{itemize}
    \item Relayer account keys
    \item Adapter owner keys
    \item User contract owner keys
    \item TWC node operation
\end{itemize}
Ideally the operator's role would be split among at least 3 entities: relay owner, Adapter owner and user contract owner. Operators are strongly encouraged to use separate keys for deploying their smart contracts and setting up the Relayer account. The exact level of risk and impact for this scenario largely depends on the deployment standards followed by the operator; a simple change in account management stops a single compromised account from cascading and compromising the entire system.  

An important aspect of this threat model is that operators are simultaneously incentivized to secure their keys and disincentized from using the TWC to performing attacks on their own contracts. This is analogous to mining pools performing attacks on their own network; doing so would simply destroy the pool's reputation and income stream. Mining pools are encouraged to act honestly and secure their operation to ensure an attacker could not gain access. Similarly, operators are encouraged to act honestly as exploiting their access to attack their own DApps would destroy their reputation and application.

An attacker will have limited capabilities in the scenario of an operator's keys becoming compromised; given a separate set of keys were used for each deployment.
\begin{itemize}
    \item Relayer account: The set of actions an attacker could perform is a subset of those an attacker could perform in compromising the TWC node. Ultimately the Adapters will identify and reject any invalid messages.
    \item Adapter owner account: Compromising the Adapter account is considered in section \ref{Adapter_compromise}.
    \item User contract: This attack vector is outside of the scope of the TWC, attacks will largely depend on functionality in the user contract.
    \item TWC Node: Compromising a TWC bridge node is considered in section \ref{Bridge_Compromise}.
\end{itemize}
The scenario of an operator's keys becoming compromised where the key is used in multiple deployments is equivalent to a multi-party compromise (considered in section \ref{Multi-Party_Compromise}).
\subsection{Multi-Party Compromise} \label{Multi-Party_Compromise}
The inherent trust model of the bridge and signatories in a notary model may result in multi-party compromise leading to catastrophic errors on the destination chain. Due to the high number of possible pairings only one of the most serious scenarios will be covered.
\subsubsection{Bridge and Signatories (Risk: Low, Impact: High)}
The destination chain relies on the bridge and signatories as its source of truth about the source chain. Given a compromised bridge and signatory set, a bridge may create any transaction; regardless of occurrence on the source chain. Compounding the issue, the compromised signatory set will sign the invalid transactions. Once the invalid transaction is transmitted to the destination chain, from the point of view of the destination contract, the transaction will be deemed valid. The destination chain will successfully validate the signatures and modify the contract state to an invalid state as specified in the falsified transaction. 

Consider a transfer of value from a compromised bridge and signatory set; attackers may mint any number of tokens on the destination chain. Once minted, these tokens are indistinguishable from real tokens. Assuming enough liquidity exists, an attacker may transfer their tokens back to the original chain and sell their newly minted tokens. An attack of this nature may modify the total supply of a token; possibly fundamentally breaking the original smart contract. This attack may only be rectified through intervention and direct modification of the smart contract state by the contract owner. 
\section{Conclusion}
Given the current state of blockchain technology, without fundamentally modifying aspects of the blockchain, there is no perfect interoperability solution. This paper presents a review of three classes of interoperability solutions; time-locks, relaying and notary schemes. Given the trade-offs between interactivity, functionality and centralization we present the TWC; a notary-based generic message passing framework for inter-blockchain communication. As a notary-based framework the TWC trades a decrease in interactivity for an increase in centralization.  

The security model provides a fault tolerant interoperability framework where a party may become compromised without compromising the entire system. However, due to the centralization trade-off, once multiple components become compromised an attacker may gain the ability to send malicious messages between blockchains. Even assuming no multi-compromise, bridge operators must still acknowledge the risk of deep chain re-organization, either by malicious actors or a coordinated fork event. Operators must be ready to step in and rectify issues caused by a deep chain re-organization deep in the chain history.   

None of the time lock, notary or relay-based frameworks provide the perfect solution for interoperability, however, the TWC presents a starting point for generic inter-blockchain message passing.

\bibliographystyle{plain}
\bibliography{paper}

\begin{thebibliography}{1}

\bibitem{buterin2016chain}
Vitalik Buterin.
\newblock Chain interoperability.
\newblock {\em R3 Research Paper}, 2016.

\bibitem{herlihy2018atomic}
Maurice Herlihy.
\newblock Atomic cross-chain swaps.
\newblock {\em arXiv preprint arXiv:1801.09515}, 2018.

\bibitem{interledger2018hashed}
Interledger.
\newblock Hashed-timelock agreements (htlas), 2018.

\bibitem{kiayias2017non}
Aggelos Kiayias, Andrew Miller, and Dionysis Zindros.
\newblock Non-interactive proofs of proof-of-work.
\newblock Technical report, Cryptology ePrint Archive, Report 2017/963, 2017.
  Accessed: 2017-10-03, 2017.

\bibitem{lerner2016drivechains}
Sergio~Demian Lerner.
\newblock Drivechains, sidechains and hybrid 2-way peg designs, 2016.

\bibitem{meckler2018coda}
Izaak Meckler and Evan Shapiro.
\newblock Coda: Decentralized cryptocurrency at scale.
\newblock 2018.

\bibitem{wood2014ethereum}
Gavin Wood.
\newblock Ethereum: a secure decentralised generalised transaction ledger.
  ethereum project yellow paper 151 (2014), 2014.

\end{thebibliography}

\newpage
\begin{appendices}
\section{Safety Conjecture for Interoperability}
We aim to informally illustrate the following: \\
\newline
\textit{Safety cannot be guaranteed for an (interoperability) system that establishes forward causality relationships between transactions on two distinct chains, where the chain which is the source for these transactions has probabilistic finality. }

\subsection{Problem Sketch}

\begin{itemize}
    \item Step I: Build a simple model for a system that establishes a forward causality relationship between transactions on two blockchains (i.e. a bridge)
    \item Step II: Precisely define a reasonable safety property of such a system. 
    \item Step III: Argue that the safety property of the interoperability system cannot be satisfied due to a dependent safety property of the source blockchain.
\end{itemize}

\subsection{Step I: Problem Modelling}

Assume the following construction for a system that establishes a \textit{forward causality} relationship between transactions on two blockchains (referred-to as a ``bridge''):
\begin{enumerate}
    \item Transaction $X_{1}$ exists on $B_{1}$ at block $N_{1}$, sealed at time \textit{T}. 
    \item $F_{1}$ is the count of blocks elapsed since $N_{1}$ at which point $X_{1}$ can be considered ``sufficiently final''.
    \item Transaction $X_{2}$ exists on $B_{2}$.
    \item $X_{2}$ has a causal relationship to $X_{1}$ (\( \hbox{i.e. $X_{1}$} \Rightarrow \hbox{$X_{2}$}\)).
    \item $X_{2}$ was sealed onto $B_{2}$ at block $ N_{1}$ + $F_{1}$, at time \textit{T} + $T_{F}$ (let $T_{F}$ = time to produce $F_{1}$ blocks on $B_{1}$) 
    \item Block production rates for $B_{1}$ and $B_{2}$ are roughly similar (impedance matching).
    \item Let $C_{1}$ be the set of claims that are invariant for blockchain $B_{1}$ at block number $N_{1}$ + $F_{1}$. (e.g. account $\varphi$ has balance of 0 at block number $N_{1}$ + $F_{1}$).
    \item Let $C_{2}$ be the set of claims that are invariant for blockchain $B_{2}$ at some time corresponding to $B_{1}$'s block number $N_{1}$ + $F_{1}$.
\end{enumerate}

\subsection{Step II: Safety Property Definition}

Our working definition of a safety property is ``[something undesirable] will not happen'' \textcolor{blue}. We define the primary safety property of our ``bridge'' construction:\\
\newline
\textit{``For all time after block $N_{1}$ + $F_{1}$, the relationship \(\hbox{$X_{1}$} \Rightarrow \hbox{$X_{2}$}\) is an invariant in the set of claims \(\hbox{$C_{1}$} \cap \hbox{$C_{2}$}\)''}\\
\newline
Another way to state this property is that the safety of our bridge system is violated if there is a non-zero possibility that the causality \(\hbox{$X_{1}$} \Rightarrow \hbox{$X_{2}$}\) established at time \textit{T} + $T_{F}$ may no-longer hold at some point in the future. 

\subsection{Step III: Proof Sketch}

For blockchains with probabilistic-finality in their consensus (e.g. Bitcoin, Ethereum), after a block is ``buried under'' sufficient proof of work, it is considered ``sufficiently final''. For probabilistic-finality blockchains, the closed-system assumption\footnote{The fact that the blockchain is a self-contained system. All security guarantees provided are reasoned about with this assumption in mind.} is central to the security model; under the closed-system assumption, blockchain reorganizations are considered perfectly safe (by definition of their safety criterion). Although never formally stated, safety criteria for probabilistic-finality blockchains\footnote{Blockchains like Bitcoin and Ethereum}, in practice, is understood to be the following:\\
\newline
\textit{Assume the set of transactions ($T_{1}$, ..., $T_{n}$), which successively depend on each other (i.e. for all n, $T_{n}$ depends on $T_{n-1}$). Assume these got sealed in the history of the blockchain at block N. }\\
\newline
\textit{Then, in the course of a re-organization to block N - M such that none of ($T_{1}$, ..., $T_{n}$) exist in the history, no guarantees are made about the re-inclusion of these transactions in the new history, other than 
if these transactions are replayed, the transaction will NOT be included out-of-order.}\\
\newline
\textit{It is perfectly reasonable that in the course of the reorganization, none or some partial contiguous series of transactions get included (starting from $T_{1}$) in the new history due to inclusion of other transactions conflicting with one or more of the transactions in the set ($T_{1}$, ..., $T_{n}$).}\\
\newline
By definition of their safety property, these systems can indefinitely tolerate a non-zero probability of a ``deep'' reorganization event. \\
\newline
Furthermore, it is easy to see from the definition of the safety criteria for a probabilistic-finality blockchain that no event $X_{1}$ sealed in a block $N_{n}$ (n \textgreater \ 0) on chain $B_{1}$ is an invariant of the chain. If the event $X_{1}$ is the source event in the forward causality relationship \(\hbox{$X_{1}$} \Rightarrow \hbox{$X_{2}$}\), it follows that this relationship also cannot be invariant (in the intersection of the claims that are invariant for all chains).
\newpage
\section{Sample Operation Using TWC}
This section outlines the steps to deploy and use the TWC adapter contracts.

Prerequisites: 
\onehalfspacing
\begin{enumerate}
    \item Deploy Adapter contracts on both Aion and Ethereum. For each contract the following should be set: 
    \begin{enumerate}
        \item \textit{\textunderscore signatoryContractAddress}: An address of a contract implementing the \textit{IBridgeSignatory} interface, storing the valid signatory information. 
        \item \textit{\textunderscore relayer}: An account address that will be used by the bridge to call the Adapter contract to process transfers. 
        \item \textit{\textunderscore signatoryQuorumSize}: Minimum number of signatures required for a transfer request to be valid. 
        \item \textit{\textunderscore transactionFee}: Fee associated with performing each cross-chain bridge transaction. 
        \item \textit{\textunderscore acceptOnlyAuthorizedSenders}: Indicates whether only a pre-authorized set of accounts can request a bridge message transfer.
    \end{enumerate}
    \item Set the source adapter contract address. Aion adapter will store the Eth adapter address and vice versa. 
    \item Encoded function in the recipient contract should be accessible by the Adapter contract. Otherwise the function call will fail. 
\end{enumerate}

Flow to send a cross-chain message transfer transaction is as follows: 
\begin{itemize}
\item On the source side of the transfer: 
\begin{enumerate}
    \item User picks the destination contract, function name and arguments.
    \item A user account or contract calls the \textit{requestTransfer} function with the destination contract address and encoded function call. Optionally the amount of gas that should be used by this function can be set as well. The sender account should be accepted in the adapter contract.  
    
    For example, an encoded function call for
    \onehalfspacing
    
    {\centering\textit{function setValue(uint128 \textunderscore v) \{ value = \textunderscore v; \} }\par}
    \onehalfspacing
   can be generated using one of the following ways: 
    \begin{enumerate}
        \item Solidity (v0.4.24): \newline
        \onehalfspacing
        \textit{abi.encodeWithSelector(bytes4(sha3(``setValue(uint128)")), 1)} 
        \item Web3: \newline
        \onehalfspacing
        \textit{ web3.eth.abi.encodeFunctionCall( \newline
        \{``name": ``setValue",  ``type": ``function",  ``inputs": [\{``type": ``uint128", ``name": ``\textunderscore v"\}] \} \newline , [1]); \ }
        
    \end{enumerate}
\end{enumerate}

Furthermore, user should include the fee set in the contract with this transaction. If the amount sent is less than \textit{transactionFee}, transaction will be rejected. Otherwise the extra value will be transferred back to the user. 

\item On the Bridge side: 

If the \textit{requestTransfer} transaction is successful, a \textit{BridgeTransferRequested} event is generated which includes an associated Id (nonce). Bridge will listen for this event and send a request to signatories asking for signatures. Signatories will validate the occurrence of the event and sign the hash of the transfer data. Hash method is Keccak256 on Ethereum and Blake2b256 on Aion. 

\textit {Hash (sourceTransactionHash, sourceAdapterAddress, recipientContract, encodedFunctionCall, gas, sourceTransferId, sourceNetworkId) }

Once the required number of signatories have signed the hash, the bridge calls the \textit{processTransfer} function on the destination adapter. 

\item On the destination side: 
\begin{enumerate}
    \item The \textit{processTransfer} function is called by the Relayer account through the bridge. 
     \item Adapter contract checks the following conditions: 
    \begin{enumerate}
    \item  Transaction hash should not have been processed before. 
    \item  Id the of the transfer request should be equal to the expected Id or nonce to maintain order between chains. 
    \item  There should be enough signatures included as the input. 
    \item  Signatures should be valid 
        
    \end{enumerate}
   \item  If the conditions pass,  
    \begin{enumerate}
    \item  The recipient contract will be called with the encoded data. 
    \item  Transaction hash will be marked as processed. 
    \item  An event will be generated indicating the success of the transaction and the status of the function call. 
    \end{enumerate}
  \item    If the transaction hash has been processed before, an event is generated which includes the block number in which the original \textit{processTransfer} transaction was sealed. This can be used to track the original transaction and ensure it's finalized.

   \item  In other fail cases, transaction will be reverted.
    \end{enumerate}
\end{itemize}
    
\begin{figure}[ht!]
  \includegraphics[width=\linewidth]{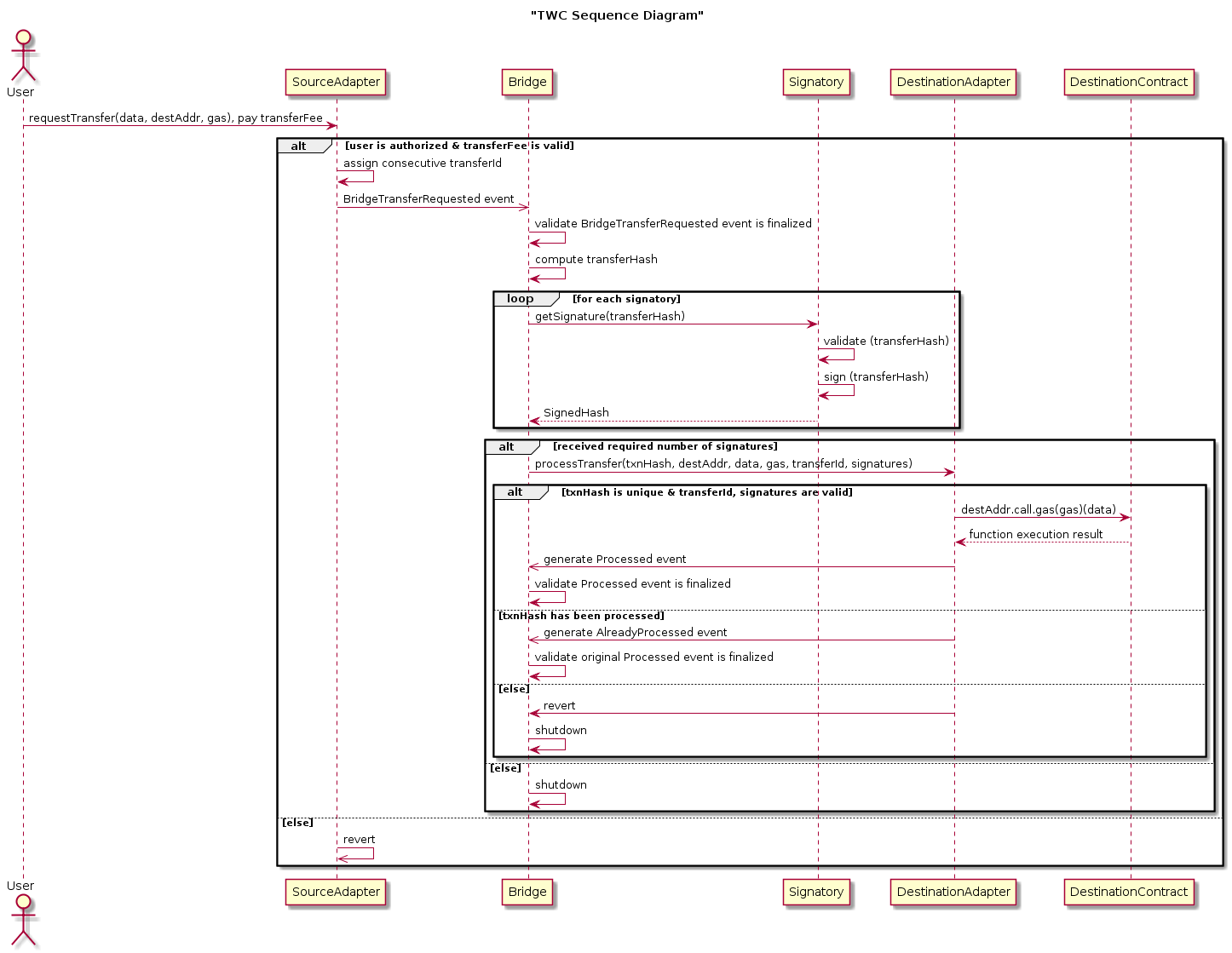}
  \caption{Sequence diagram of the TWC protocol}
  \label{Figure 2}
\end{figure}
\newpage
\section{Contract code} \label{contract_code}
\textit{We do not recommend using these contracts in a production environment. The code has not been fully tested or audited.} Smart contracts for both Aion and Ethereum can be found in 
\url{https://github.com/aionnetwork/transwarp_conduit}

\end{appendices}

\end{document}